\documentclass[11pt, a4paper]{article}
\pdfoutput=1

\usepackage[english]{babel}
\usepackage[utf8]{inputenc}
\usepackage{jcappub_ver}

\newcommand{\jvem}{}
\newcommand{\bea}{\begin{eqnarray}}
\newcommand{\eea}{\end{eqnarray}}

\newcommand{\pat}{\partial}
\renewcommand{\sec}[1]{section \ref{#1}}
\newcommand{\fig}[1]{figure \ref{#1}}
\newcommand{\tab}[1]{table \ref{#1}}
\newcommand{\para}{\paragraph}

\renewcommand{\a}{\alpha}
\renewcommand{\b}{\beta}
\renewcommand{\c}{\gamma}
\renewcommand{\d}{\delta}
\newcommand{\e}{\epsilon}
\newcommand{\LCDM}{$\Lambda$CDM\ }
\newcommand{\rmd}{\mathrm{d}}

\newcommand{\nonum}{\\}

\newcommand{\ie}{i.e.\ }

\newcommand{\adot}{\dot{a}}

\newcommand{\dr}{\mathrm{dr}}

\newcommand{\half}{\frac{1}{2}}

\newcommand{\Om}{\Omega_{\mathrm{m}}}
\newcommand{\om}{\omega_{\mathrm{m}}}

\newcommand{\ob}{\omega_{\mathrm{b}}}

\newcommand{\oc}{\omega_{\mathrm{c}}}

\newcommand{\Neff}{N_{\mathrm{eff}}}

\newcommand{\APJ}[1]{{Astrophys. J.} {\bf #1}}
\newcommand{\GRG}[1]{{Gen. Rel. Grav.} {\bf #1}}

\title{Testing distance duality with CMB anisotropies}

\author{Syksy R\"{a}s\"{a}nen,}

\author{Jussi V\"{a}liviita}

\author{and Ville Kosonen}

\affiliation{University of Helsinki, Department of Physics
and Helsinki Institute of Physics,\\
P.O. Box 64, FIN-00014 University of Helsinki, Finland}

\emailAdd{syksy.rasanen@iki.fi}
\emailAdd{jussi.valiviita@helsinki.fi}
\emailAdd{ville.kosonen@helsinki.fi}

\abstract{
We constrain deviations of the form $T\propto (1+z)^{1+\epsilon}$ from the
standard redshift-temperature relation, corresponding to modifying distance
duality as $D_L=(1+z)^{2(1+\epsilon)} D_A$. We consider a consistent model,
in which both the background and perturbation equations are changed. For
this purpose, we introduce a species of dark radiation particles to which
photon energy density is transferred, and assume $\epsilon\ge0$. The Planck
2015 release high multipole temperature plus low multipole data give the
limit $\epsilon<4.5\times 10^{-3}$ at 95\% C.L. The main obstacle to
improving this CMB-only result is strong degeneracy between $\epsilon$ and
the physical matter densities $\omega_{\rm b}$ and $\omega_{\rm c}$. A
constraint on deuterium abundance improves the limit to $\epsilon<1.8\times
10^{-3}$. Adding the Planck high-multipole CMB polarisation and BAO data
leads to a small improvement; with this maximal dataset we obtain
$\epsilon<1.3\times 10^{-3}$. This dataset constrains the present
dark radiation energy density to at most 12\% of the total photon plus dark
radiation density. Finally, we discuss the degeneracy between dark radiation
and the effective number of relativistic species $N_{\rm eff}$, and consider the
impact of dark radiation perturbations and allowing $\epsilon<0$ on the results.
}

\keywords{CMBR theory, cosmological parameters from CMBR, cosmological perturbation theory, cosmological neutrinos}

\begin{document}

\begin{flushleft}
	\hfill		 HIP-2015-41/TH\\
        \hfill                 Published in JCAP {\bf 04} (2016) 050
\end{flushleft}

\maketitle
  
\setcounter{tocdepth}{2}

\setcounter{secnumdepth}{3}

\section{Introduction} \label{sec:intro}

\para{Temperature-redshift relation and distance duality.}

The proportionality $T\propto(1+z)$ between the temperature $T$ of
massless particles with a thermal phase space 
distribution and the redshift $z$
has a central role in the history of the universe.
This relation is not restricted to the exactly homogeneous and isotropic
Friedmann--Robertson--Walker (FRW) models. It holds
in general relativity (and similar theories with massless particles minimally
coupled to the metric) as long as particle number is conserved
and the geometrical optics approximation holds, \ie the spacetime curvature scale is much larger than the wavelength and the wavefront curvature radius.
In strong fields, where the spacetime curvature scale is $\gtrsim T$, the geometrical optics approximation is not valid, massless particles do not travel on null geodesics, and the relation $T\propto(1+z)$ does not in general hold.
For cosmological observations, the geometrical optics approximation
is well satisfied, so any deviation from $T\propto(1+z)$ would be a
clear signal for new fundamental physics.
Proposed possibilities include decaying vacuum energy, coupling of photons to  exotic light particles and disformal coupling of photons to a scalar field \cite{Freese:1987, models, Avgoustidis, Jetzer:2012}.

Soon after the discovery of the cosmic microwave background (CMB),
it was realised that the CMB excites atoms and molecules in the
intergalactic medium, so absorption line measurements of the ratio of ground
and excited levels can be used to find $T$ as a function of $z$ \cite{Bahcall:1968}.
Such observations have been made up to $z=3.025$ \cite{absor, Muller:2012}.
Parametrising the temperature as $T\propto(1+z)^{1+\e}$ \cite{Lima:2000},
the latest constraints are $\e=-0.009\pm0.019$ ($-0.047<\e<0.029$) \cite{Muller:2012} and $\e=-0.005\pm0.022$ ($-0.049<\e<0.039$) \cite{Avgoustidis:2015}
(we quote error bars as the 68\% C.L. region and limits as the 95\% C.L. region).
Another way to determine $T(z)$ is to observe the spectral distortion
due to the thermal Sunyaev--Zel'dovich (tSZ) effect, \ie collisions
of CMB photons with hot electrons.
Such measurements have been made up to redshift $z=1.35$, and they give $\e$ consistent with zero, with 95\% C.L. ranges comparable to those from absorption data \cite{tSZ, Avgoustidis:2011, Jetzer:2012}. Latest combination of absorption and tSZ data gives $\e=-0.0064\pm0.0086$ ($-0.024<\e<0.011$) \cite{Avgoustidis:2015}.

Violation of the temperature-redshift relation is tied to violation of the distance duality relation
$D_L = (1+z)^2 D_A$ between the luminosity distance $D_L$
and the angular diameter distance $D_A$ \cite{Etherington:1933, Ellis:1971}.
Distance duality has been observationally tested both using the relation between $T$ and $z$, as well as using a direct comparison of the distances \cite{duality, Avgoustidis, Avgoustidis:2015}.

In addition to looking at the tSZ effect, the CMB has been used by the Planck collaboration to constrain the $T(z)$ relation by looking at the effect of change in the CMB monopole on the CMB anisotropies \cite{Planck:cosmo}. The resulting limit is $\e=(-0.2\pm1.4)\times10^{-3}$ or $\e=(-0.4\pm1.1)\times10^{-3}$, depending on the dataset (we discuss this in \sec{sec:Planck}).
Only a change in the background evolution of $T(z)$ was considered.
However, any new physics that leads to violation of the background
scaling $T\propto(1+z)$ will in general also affect perturbations directly.
In general, models that have identical background evolution can differ
at the level of perturbations, so a consistent perturbative analysis
will be more model-dependent. 
By the same token, perturbations have to be included in the analysis in order to evaluate how much neglecting them affects the results.

We consider one of the simplest possibilities of upgrading the relation $T\propto (1+z)^{1+\e}$ to the perturbative level consistently, by assuming that the change in scaling is due to photon interaction with dark radiation (dr) particles. Our main results for $1000\e$ are presented on the first line of \tab{tab:numbers} on page \pageref{tab:numbers}. We also compare the results to the (unphysical) case where only the background change is taken into account.

\section{Setting up the model} \label{sec:model}

\subsection{Distance duality and the CMB temperature} \label{sec:dua}

\para{Distance duality.}

The relation between the temperature-redshift relation and
distance duality can be appreciated by considering how we could
use the CMB to measure the luminosity distance, defined as
\bea
  D_L = \sqrt{\frac{L}{4\pi F}} \,,
\eea
where $L$ is source luminosity and $F$ is the observed flux.
For a source with blackbody spectrum at temperature $T$,
we have $L\propto A T_s^4$ and $F\propto T_o^4$, where $A$ is the
area of emission and $s$ refers to source and $o$ to observer.
Part of the CMB sky with angular size $\rmd\Omega$ has
the area $\rmd A=D_A^2\rmd\Omega$, where $D_A$ is the angular
diameter distance. On the other hand, the flux from that part
of the sky is $\rmd\Omega/(4\pi)$ of the total
(neglecting the small anisotropies in the flux),
so we have
\bea \label{duaT}
  D_L = \left( \frac{T_s}{T_o} \right)^2 D_A \,.
\eea
Therefore, the CMB monopole cannot be used to measure
the luminosity distance, unless the angular diameter distance
is independently known. The reason is that, in contrast to localised sources like supernovae, the CMB sources form a continuum, and their emission cannot be disentangled from each other.

For photons travelling along null geodesics in general relativity, with conserved photon number, we have $T\propto (1+z)$, so \eqref{duaT} is equivalent to the distance duality relation, which follows from the Etherington reciprocity relation \cite{Etherington:1933, Ellis:1971}
\bea \label{ether}
  D_L = (1+z)^2 D_A \,.
\eea
Thus, the fact that the CMB monopole cannot be used to independently measure the luminosity distance can be turned around to say that testing the relation $T\propto (1+z)$ also tests distance duality, assuming that deviations from the blackbody shape can be neglected (significant scattering by dust would change the relation, for example). Possible violation is often parametrised as $T\propto (1+z)^{1+\e}$ \cite{Lima:2000}, corresponding to
\bea \label{duaz}
  D_L = (1+z)^{2(1+\e)} D_A \,,
\eea
with $\e$ taken to be constant for simplicity.\footnote{It
is common to use $-\e$ in the exponent instead of $\e$. As we
focus on the case where the temperature increases faster with redshift
than usual, we prefer not to have the minus sign, so that $\e\ge0$.}
We consider this simple ansatz.

\para{Violating the distance duality.}

Violating distance duality means abandoning travel on null
geodesics of the metric or photon number conservation. Either
will, in general, lead to deviations from the blackbody
spectrum \cite{Avgoustidis:2011, Ellis:2013, Chluba:2014}.
In the usual case, photon energy scales as $E\propto(1+z)$, photon number density goes as $n_\c\propto T^3$, energy density as $\rho_\gamma=\langle E\rangle n_\gamma\propto T^4$ (where $\langle\rangle$ is the phase space average), and the temperature scales as $T\propto(1+z)$.

Let us first assume that photon energy scales as $E\propto(1+z)$ as usual, but the temperature as $T\propto(1+z)^{1+\e}$, where $T$ is defined by the relation $\rho_\c\propto T^4$, \ie $T$ is the bolometric temperature. The number density then evolves as \mbox{$n_\c\propto(1+z)^{3+4\e}\simeq(1+z)^3 [1+4\e\ln(1+z)]$}, where we have assumed $|\e|\ll1$. Deviations from the blackbody shape are usually parametrised in terms of the $y$-distortion \cite{Sunyaev:1970}, which corresponds to a certain reshuffling of photon energies with the photon number kept constant, and the $\mu$-distortion \cite{Zeldovich:1969}, which corresponds a non-zero chemical potential.
The respective amplitudes are constrained as $|y|<1.5\times10^{-5}$ and $|\mu|<0.9\times10^{-4}$ \cite{Fixsen:1996}. These are not a complete parametrisation of all possible distortions \cite{Pitrou:2009, Khatri:2012, Chluba:2013}, but the constraints on other distortions are expected to be similar, giving
$|\e|\ln(1+z_i)\lesssim10^{-4}$, where $z_i$ is the redshift at which the violation of the scaling $T\propto(1+z)$ turns on. Except in the extreme case $z_i\ll1$, we have $|\e|\lesssim10^{-4}$.
Note that even if we only multiplied the photon occupation number by a factor which is independent of energy, the spectrum would be distorted by a greybody factor, which is constrained to be $\lesssim10^{-4}$ \cite{Ellis:2013}.
If we instead change the scaling of the energy to $E\propto(1+z)^{1+4\e}$, but keep the number density scaling as $n_\c\propto T^3$, \ie $T$ is the occupation number temperature, and $T\propto(1+z)$, we get a generalised $y$-distortion \cite{Pitrou:2009, Khatri:2012, Chluba:2013}, and similar limits as above.

Violations of the temperature-redshift scaling, and therefore distance duality, are thus strongly constrained, unless they change both the scaling of the energy density and the number density, without affecting the distribution function. Such change has to be fine-tuned: photon injection or removal must be adiabatic \cite{Avgoustidis:2011} and satisfy a strict spectral condition \cite{Chluba:2014} (see also \cite{Chluba:2015}).

In the analyses of the CMB temperature using spectral lines and the tSZ effect, restricted to redshifts $z\leq3.025$, it has been sufficient to consider the change in the background relation $T\propto(1+z)$, without specifying the physics responsible for the change of the scaling. This approach was also adopted in the analysis based on CMB anisotropies, reaching $z=1090$ \cite{Planck:cosmo}.
We will instead consistently consider changes also in the perturbation equations.
We introduce a gas of dark radiation particles with a coupling to
photons chosen so as to obtain $T\propto(1+z)^{1+\e}$, although
we will not specify the microphysics of the interactions.

\subsection{Mixing of photons and dark radiation \label{sec:mixing}}

\para{Motivation.}

We want to reproduce the scaling $T\propto(1+z)^{1+\e}$ for the
background temperature in a consistent model that includes perturbations.
In terms of the background photon energy density
$\bar\rho_\c\propto T^4\propto (1+z)^{4(1+\e)}\propto a^{-4(1+\e)}$,
we have
\bea \label{rhogamma1}
  \dot{\bar\rho}_\c + 4 H \bar\rho_\c &=& - \e\, 4 H \bar\rho_\c \,,
\eea
where overbar refers to background quantities, $a$ is the scale factor, dot refers to derivative with respect to conformal time $\eta$, and $H\equiv\adot/a$ is the Hubble parameter defined with respect to $\eta$. A simple way to keep \eqref{rhogamma1} consistent with the covariant conservation of the energy-momentum tensor is to add a new energy density component into which photon energy density is transferred. We therefore introduce a gas of dark radiation particles, with energy density $\bar\rho_\dr$ that satisfies
\bea \label{rhodr1}
  \dot{\bar\rho}_\dr + 4 H \bar\rho_\dr &=& \e\, 4 H \bar\rho_\c \,,
\eea
so that
$\dot{\bar\rho}_\c + \dot{\bar\rho}_\dr + 4 H (\bar\rho_\c+\bar\rho_\dr)=0$.

These equations can be promoted to the perturbative level in different
ways. We generalise \eqref{rhogamma1} by demanding that it holds also
for the local photon energy and the local expansion rate, as measured
in the photon gas rest frame. We assume that the dr has no
non-gravitational interactions with any particles other than photons.
We also assume that the interaction between the photon gas and dr does not have  preferred spatial directions in the photon gas rest frame (so there is no net momentum transfer between the gases). We now show that these assumptions uniquely determine the equations for the dr and photons.

\para{Exact treatment.}

Because dr interacts only with photons, we have, neglecting here the interactions between photons and ordinary matter (see e.g. \cite{Malik:2002}; for the covariant formalism, see e.g. \cite{cov}),
\bea 
  \label{mix1} \nabla_\a T^\a_{(\c) \b} = - Q_\b \\
  \label{mix2} \nabla_\a T^\a_{(\dr) \b} = Q_\b \,,
\eea
where $Q_\b$ is the energy-momentum transfer vector. (When dealing with the exact equations, we enclose the species label in parentheses so as not to confuse the photon symbol $\c$ with spacetime indices.)

Calculating the CMB anisotropies involves following the Boltzmann hierarchy, and with current precision data, it is important to take into account the anisotropic stress \cite{Sellentin:2014}. Allowing for anisotropic stress (but not energy flux), the energy-momentum tensors are
\bea \label{T}
  T_{(n)\a\b} = ( \rho_{(n)} + p_{(n)} ) u_{(n)\a} u_{(n)\b} + p_{(n)} g_{\a\b} + \pi_{(n)\a\b} \,,
\eea
where $n=\c,\dr$ and $\rho_{(n)}$, $p_{(n)}$, $u_{(n)\a}$,
$g_{\a\b}$ and $\pi_{(n)\a\b}$ are the energy density, pressure,
four-velocity, metric, and anisotropic stress, respectively.
Both photons and dr are ultrarelativistic (perhaps exactly massless), so $p_{(n)}=\frac{1}{3}\rho_{(n)}$.

From our assumption that, in the photon gas rest frame, there are no preferred
spatial directions for the interaction between photons and dr, it follows that $Q_\a=Q u_{(\c)\a}$. For example, this form would describe photon decay. Combining \eqref{mix1} and \eqref{T} and projecting with $u_{(\c)}^{\a}$ gives
\bea \label{Q1}
  Q = - u_{(\c)}^\a \nabla_\a \rho_{(\c)} - \frac{4}{3} \rho_{(\c)} \nabla_\a u_{(\c)}^\a - \sigma_{(\c)\a\b} \pi_{(\c)}^{\a\b} \,,
\eea
where $\sigma_{(\c)\a\b}$ is the photon shear tensor.
Generalising the background equation \eqref{rhogamma1} to hold also for
the corresponding local quantities, we have
\bea
  u_{(\c)}^\a \nabla_\a \rho_{(\c)} + \frac{4}{3} ( 1 + \e ) \rho_{(\c)} \nabla_\a u_{(\c)}^\a = 0 \,,
\eea
which in combination with \eqref{Q1} fixes the energy-momentum
transfer uniquely to be
\bea \label{Q2}
  Q = \frac{4}{3} \e \rho_{(\c)} \nabla_\a u_{(\c)}^\a - \sigma_{(\c)\a\b} \pi_{(\c)}^{\a\b} \,.
\eea
This transfer rate between photons and dr
is unusual in that it depends on the local expansion rate
$\nabla_\a u_{(\c)}^\a$; this is of course already the case
for the background equation \eqref{rhogamma1}.
We do not specify the microphysics leading to \eqref{Q2}.
But as the transfer has been defined with respect to the photon frame,
it seems difficult to justify $\e<0$, as the energy transfer from dr to photons would depend on the photon energy density.
We therefore focus on the case $\e\geq0$, so that photons are removed from the thermal bath, not added, though mathematically the equations are well-defined also for $\e<0$.
For completeness, in \sec{sec:Planck} we repeat our analysis with $\e<0$ not excluded, which also allows closer comparison to the Planck analysis \cite{Planck:cosmo}.
As discussed above, the interaction has to be specifically tuned to avoid distortions of the blackbody spectrum larger than $10^{-4}$, though this may be less difficult to engineer for photon removal than in the case of photon injection \cite{Chluba:2011, Khatri:2012, Chluba:2013, Chluba:2014, Chluba:2015}.

\para{Background.}

We consider first order perturbation theory around a spatially
flat FRW universe in the synchronous gauge,
\bea
  \rmd s^2 = a(\eta)^2 [ - \rmd\eta^2 + ( \d_{ij} + h_{ij} ) \rmd x^i \rmd x^j ] \,.
\eea
For the background, the energy continuity equations \eqref{mix1} and \eqref{mix2} with \eqref{Q2} reproduce the desired scaling laws,
\bea
  \label{rhodotgamma} \dot{\bar\rho}_\c + 4 ( 1 + \e ) H \bar\rho_\c &=& 0 \\
  \label{rhodotdr} \dot{\bar\rho}_\dr + 4 H \bar\rho_\dr &=& 4 \e H \bar\rho_\c \,.
\eea
Because we have $\bar\rho_\c+\bar\rho_\dr\propto a^{-4}$, the evolution of the background scale factor can remain unchanged even when there are significant amounts of dr. Thus, the limits on relativistic extra degrees of freedom \cite{Rossi:2014} do not straightforwardly apply. However, because $\bar\rho_{\c0}=\textstyle \frac{\pi^2}{15} \bar T_0^4$ is fixed by the observed $\bar{T}_0$ \cite{Fixsen:2009ug}, the presence of any dr today implies larger radiation density in the past, and $\e$ will in fact turn out to be correlated with the number of other relativistic degrees of freedom $\Neff$, as we discuss in \sec{sec:Neff}.

In addition to $\e$, we need one parameter that sets the redshift $z_i$ at which the scaling $T\propto (1+z)^{1+\e}$ starts to apply. We have
$\bar\rho_\dr = ( \bar\rho_{\dr0} + \bar\rho_{\c0} ) a^{-4} - \bar\rho_{\c0} a^{-4(1+\e)}$,
where the subscript $0$ refers to the present day, with $a_0=1$.
Therefore we would have $\bar\rho_\dr<0$ at some point in the past if the scaling extended arbitrarily far backwards. As the new parameter, which (given $\e$) determines the present dr energy density $\bar\rho_{\dr0}$ as well as $z_i$, we choose the fractional contribution of dr at $z_{\mathrm{ref}}\equiv1090$, roughly corresponding to last scattering,
\bea
  f_{\dr*} \equiv \frac{\bar\rho_{\dr}(z_{\mathrm{ref}})}{\bar\rho_{\c}(z_{\mathrm{ref}}) + \bar\rho_{\dr}(z_{\mathrm{ref}})} \,,
\label{eq:fdr}
\eea
and we define a derived parameter $f_{\dr0}$ in the same way, with the densities evaluated today.
The present day dr density can be written as
\bea
\bar\rho_{\dr0} = \left[ \frac{1}{(1- f_{\dr*}) a_{{\mathrm{ref}}}^{4\e}} - 1 \right] \bar\rho_{\c0}\,,
\label{eq:rhodr0}
\eea
where $a_{\mathrm{ref}}=(1+z_{\mathrm{ref}})^{-1}=1/1091$.
Solving the scale factor $a_i$ from $\bar\rho_\dr(a_i)=0$ we find
\bea
a_i = a_{\mathrm{ref}} \left( 1- f_{\dr*} \right)^{1/(4\e)}\,,
\eea
and the redshift at which the interaction between photons and dr turns on both in the background and perturbation equations is $z_i = 1/a_i -1$. Before this time we keep $\bar\rho_\dr$ and $\e$, as well as all multipoles of the dr perturbation hierarchy, equal to zero.

Of course, the choice of $ f_{\dr*}$ as a primary dr parameter is not unique. We could equally well have defined $f_{\dr}$ at some other redshift (far enough in the past), or $\bar\rho_{\dr}$ at a given redshift, or even used directly $z_i$. Different primary parameters lead to different integration measures upon marginalisation, affecting the end results to some extent. In order to test the robustness of our results, we repeated parts of our analysis by using $-\log_{10}z_i$ as a primary parameter, with a uniform prior. (We include the minus sign in the plots, as we prefer time running from left to right.) This led to slightly tighter upper bounds on $\e$ and $f_{\dr*}$, than using $f_{\dr*}$ as a primary parameter.  To be conservative, we report the results from the analysis where $f_{\dr*}$ is a primary parameter.

Care should be taken about the choice of the primary parameter, so that the implied prior does not dominate the posterior. Our choice guarantees that the photon-dr interaction is turned on at least for the whole time between last scattering and today, \ie $z_i > z_{\mathrm{ref}}\approx z_*$, where $z_*$ is the redshift of last scattering. The closer to zero $f_{\dr*}$ is, the less time before last scattering the interaction is turned on. For a fixed positive $f_{\dr*}$, we have $\e\rightarrow \infty$ $\Rightarrow$ $z_i \rightarrow z_{\mathrm{ref}}$. This kind of behaviour would cause a problem if we chose e.g. $f_{\dr0}$ as the primary parameter, \ie $z_{\mathrm{ref}}=0$.
If a non-zero interaction is not preferred by the data (as is the case), it is favourable to minimise the impact of dr by making the time that it disturbs the standard evolution as short as possible.
For a fixed positive $f_{\dr0}$, this would be achieved by an arbitrary large $\e$, since this would cause $z_i \rightarrow 0$ and thus eliminate the interaction altogether. This would bias the results toward large positive values of $\e$.

As we use CMB anisotropies to constrain the interaction, we concentrate on the case where the interaction is present at least from last scattering to the present day, hence the choice \eqref{eq:fdr}. This parametrisation allows the range $1090 < z_i < \infty$. However, it would be unphysical to start the modified photon temperature scaling before reheating, \ie the earliest possible time that the radiation can have been produced. Hence, we apply a ``top-hat'' prior $z_i<10^{29}$, set by the maximum reheating temperature. The limit comes from the constraint on the energy scale of inflation due to the upper bound $r<0.07$ on the tensor-to-scalar ratio \cite{r}. (The limit depends on the number of degrees of freedom at reheating only very weakly, $z_i\propto g_*(T_{\mathrm{rh}})^{\frac{1}{12}}$.) We will see in \fig{fig:dr_1d} and other figures that the prior $-\log_{10}z_i>-29$ excludes only a small fraction of models, as the data prefer the interaction to turn on later. 

\para{Perturbations.}

We have, in the notation of \cite{Ma:1995}, to first order in perturbations,
\bea \label{theta}
  \nabla_\a u_{(n)}^\a = 3 H + \theta_n + \frac{1}{2} \dot h \,,
\eea
where $\theta_n\equiv a \pat_i u_{(n)}^i$, $h\equiv\delta^{ij} h_{ij}$. Inputting \eqref{theta} into \eqref{mix1} and \eqref{Q2} (the last term in \eqref{Q2} does not contribute at first order) and using \eqref{T}, and the fact that $u_{(n)}^0= -a^{-2} u_{(n)0}=a^{-1}$ to first order, we get
\bea \label{deltac}
  0 = \dot\delta_\c + \frac{4}{3} ( 1 + \e ) \theta_\c + \frac{2}{3} ( 1 + \e ) \dot h \,.
\eea
This replaces the first line of equation~(63) in \cite{Ma:1995}. Other perturbation equations for photons remain unchanged.

In addition to including \eqref{deltac} and the altered background, we modify the Code for Anisotropies in the Microwave Background (\texttt{CAMB}, \cite{CAMB}) to include the Boltzmann hierarchy for the perturbation evolution of a new species of massless radiation. It is identical to that of massless neutrinos, except that the evolution of the density contrast $\delta_\dr$ (the zeroth multipole) is given by
\bea \label{deltadr}
  \dot\delta_\dr + \frac{4}{3} \theta_\dr + \frac{2}{3} \dot h = \e \frac{\bar\rho_\c}{\bar\rho_\dr} \left( \frac{4}{3} \theta_\c + \frac{2}{3} \dot h + 4 H \delta_\c - 4 H \delta_\dr \right)\,,
\eea
which replaces the first line of equation (49) in \cite{Ma:1995}. For the general derivation of perturbation equations of interacting fluids in an arbitrary gauge, see \cite{Kodama:1985bj, ide}; \eqref{deltac} and \eqref{deltadr} can be straightforwardly read from \cite{ide}.

We also consider two alternative treatments of the dr perturbations. In one, we truncate their Boltzmann hierarchy at the ideal fluid level, \ie we set anisotropic stress and all higher multipoles to zero, as done in \cite{Sellentin:2014} for neutrinos. We also study the case where the dr perturbations are zero (in the synchronous gauge); this involves setting $\e=0$ in \eqref{deltac} to keep the energy-momentum tensor covariantly conserved at the perturbative level. Neglecting perturbations is not physically meaningful, but it allows us to compare to the Planck analysis \cite{Planck:cosmo}, where only change in the background scaling $T\propto(1+z)$ was considered.

As we assume that there is no dr initially, at the moment $t_i$ when the interaction between photons and dr turns on, we have $\bar\rho_\dr(t_i)=0$. The initial condition for the dr density contrast follows from \eqref{deltadr}, using \eqref{rhodotdr}; at $t=t_i$, we have
\bea \label{drinit}
  \delta_\dr = \delta_\c + \frac{1}{3 H} \left( \theta_\c + \frac{1}{2} \dot h \right) \,.
\eea

\setlength{\tabcolsep}{.33667em}
\begin{table}[t]
\footnotesize
\centering
\begin{tabular}{l|rrrr|rrrr}
\hline
\hline
                                        &  \multicolumn{4}{l|}{TT+lowP} & \multicolumn{4}{l}{TT,TE,EE+lowP}  \\
                                        &     & +D/H    &  +BAO &  +BAO&        & +D/H        & +BAO           & +BAO  \\
                                        &      &       &         &  +D/H &                 &        &                  & +D/H  \\
\hline
$1000\epsilon\!:\!\ \ $ full dr &   4.5 &   1.8 &   4.4 &   1.6 &   2.1 &   1.5 &   2.0 &   1.3  \\
$\phantom{1000\epsilon\!\!:\ \ }$ $\Lambda$CDM parameters fixed &   0.4 &   0.4 & & & & &   0.3 &   0.3  \\
$\phantom{1000\epsilon\!\!:\ \ }$ + nuisance parameters fixed &   0.3 &   0.3 & & & & &   0.2 &   0.2  \\
$\phantom{1000\epsilon\!\!:\ \ }$  full dr + $\Neff$ &   6.6 &   4.7 & & & & &   2.8 &   2.5  \\
$\phantom{1000\epsilon\!\!:\ \ }$ ideal fluid dr &   4.6 &   1.9 & & & & &   1.9 &   1.3  \\
$\phantom{1000\epsilon\!\!:\ \ }$ no dr perturbations &   4.2 &   2.5 & & & & &   3.4 &   2.1  \\
\hline
$100f_{\rm dr\ast}\!\!:$ full dr &  25.9 &  12.8 &  23.7 &  13.0 &  12.2 &   9.2 &  12.7 &   9.5  \\
$\phantom{1000\epsilon\!\!:\ \ }$ $\Lambda$CDM parameters fixed &   1.5 &   1.5 & & & & &   0.9 &   0.9  \\
$\phantom{1000\epsilon\!\!:\ \ }$ + nuisance parameters fixed &   0.9 &   0.9 & & & & &   0.6 &   0.6  \\
$\phantom{1000\epsilon\!\!:\ \ }$  full dr + $\Neff$ &  32.5 &  12.3 & & & & &  18.6 &   9.2  \\
$\phantom{1000\epsilon\!\!:\ \ }$ ideal fluid dr &  18.1 &   7.8 & & & & &   9.6 &   7.7  \\
$\phantom{1000\epsilon\!\!:\ \ }$ no dr perturbations &   5.7 &   5.0 & & & & &   4.4 &   3.6  \\
\hline
$100f_{\rm dr0}\!\!:$ full dr &  34.5 &  16.1 &  32.6 &  16.2 &  16.2 &  12.0 &  16.7 &  12.1  \\
$\phantom{1000\epsilon\!\!:\ \ }$ $\Lambda$CDM parameters fixed &   2.3 &   2.3 & & & & &   1.4 &   1.4  \\
$\phantom{1000\epsilon\!\!:\ \ }$ + nuisance parameters fixed &   1.5 &   1.5 & & & & &   1.0 &   1.0  \\
$\phantom{1000\epsilon\!\!:\ \ }$  full dr + $\Neff$ &  40.4 &  19.4 & & & & &  22.1 &  12.5  \\
$\phantom{1000\epsilon\!\!:\ \ }$ ideal fluid dr &  26.7 &  10.8 & & & & &  13.3 &  10.1  \\
$\phantom{1000\epsilon\!\!:\ \ }$ no dr perturbations &  15.9 &  11.0 & & & & &  13.2 &   8.8  \\
\hline
\hline
\end{tabular}
\caption{\label{tab:numbers}The 95\% C.L. upper bounds on the dark radiation parameters with various datasets.
For each parameter, ``full dr'' gives the result for our baseline case, where the full Boltzmann hierarchy for dr is considered, and in addition to $\epsilon$ and $f_{\rm dr\ast}$ also all the other 6 cosmological parameters ($\ob$, $\oc$, $\theta_\ast$, $\tau$, $\log A_s$, $n_s$) as well as the nuisance/foreground parameters of the Planck likelihoods (15 for TT+lowP and 27 for TT,TE,EE+lowP) are varied. Then, in the ``\LCDM parameters fixed'' case, the above-mentioned 6 cosmological parameters are kept at their best-fit vanilla \LCDM values. In the ``+ nuisance parameters fixed'' case, the nuisance/foreground parameters of the Planck likelihoods are in addition fixed to their best-fit vanilla \LCDM values. The case ``full dr + $\Neff$'' is the same as ``full dr'', but the number of neutrino degrees of freedom is extended from 3.046 to the free parameter $\Neff$. In the ``ideal fluid dr'' case, dr is treated as an ideal fluid, \ie only the zeroth (density contrast, $\delta_\dr$) and the first multipole (fluid velocity divergence, $\theta_\dr$) of the Boltzmann hierarchy are kept. In the ``no dr perturbations'' case, the dr perturbations are turned off, so only the background evolution equations are modified with respect to the vanilla \LCDM case.}
\end{table}

\section{Data and methods}

We use CMB data from the Planck 2015 data release \cite{Adam:2015rua,Aghanim:2015xee} available via the Planck Legacy Archive (PLA), supplemented by
baryon acoustic oscillation (BAO) data\footnote{The BAO data analysis leading to the published values has been done assuming the standard distance duality relation. However, this makes no difference for our results, because the constraints we obtain without BAO data on the violation of the distance duality relation are tighter than the precision of the BAO constraints, and adding the BAO data has little impact on our results.}
from the 6dFGS, BOSS and SDSS DR7 surveys \cite{BAO}, labelled 6DFGS, DR11CMASS, DR11LOWZ and SDSS MGS in \cite{Planck:cosmo}, and a Gaussian prior on the deuterium abundance, D/H=$(2.53\pm0.04)\times10^{-5}$, from Lyman-$\alpha$ absorption lines \cite{Cooke:2013}. We consider the Planck data in two different sets. The first consists of the high multipole temperature anisotropy data (TT), named in PLA as \texttt{plik\_dx11dr2\_HM\_v18\_TT.clik}, plus the low multipole (large scale) temperature, E polarisation and B polarisation data (lowP, \texttt{lowl\_SMW\_70\_dx11d\_2014\_10\_03\_v5c\_Ap.clik}, also dubbed \texttt{lowTEB}). The second dataset contains, in addition to the previous ones, the high-multipole E polarisation data (TE+EE), called in PLA \texttt{plik\_dx11dr2\_HM\_v18\_TTTEEE.clik}.
In all figures dashed lines refer to the first dataset (possibly with non-CMB data added), and solid lines or shaded colours refer to the case where also high-multipole polarisation data are utilised.

Our baseline dataset is Planck TT+lowP, and unless otherwise stated, our quoted constraints refer to it. The maximal dataset is TT,TE,EE+lowP+BAO+D/H. As indicated in \tab{tab:numbers}, we find the posterior probability densities for 8 different data combinations in our baseline full dr model. We study also five special cases with 4 data combinations. We scan the likelihood surface using Markov Chain Monte Carlo (MCMC) method as implemented in \texttt{CosmoMC} \cite{CosmoMC}, which calls our modified version of \texttt{CAMB} \cite{CAMB} to calculate the theoretical temperature and polarisation angular power spectra for each set of the values of parameters drawn by \texttt{CosmoMC}. In addition to the changes discussed above, we have modified the calculation of the sound horizon to be consistent with the dr model; see appendix \ref{sec:standard} for details. The number of our \texttt{CosmoMC} runs for the dr model(s), each containing 8 Markov Chains, is 14. The data combinations containing D/H are importance sampled from the other runs. For comparison we also perform MCMC analysis of the vanilla \LCDM model for the 8 data combinations, and the $\Lambda$CDM+$\Neff$ case for 2 data combinations. 
Finally, we rerun 8 cases allowing also for negative values of $\epsilon$.

We fix the number of (other than photon and dr) relativistic degrees of freedom, corresponding to the  three neutrino species, to $\Neff=3.046$, except in the runs described in \sec{sec:Neff} where we leave $\Neff$ as a free parameter. Following the convention established by the Planck collaboration, we assume two massless neutrinos and one massive, and set $\sum m_\nu = 0.06\,$eV.

\section{Results}

As discussed near the end of \sec{sec:mixing}, we consider three different treatments of the dr perturbations: (1) full perturbation equations  with the usual Boltzmann hierarchy, (2) assuming dr to be an ideal fluid and (3) setting the dr perturbations to zero (in this case, to guarantee consistent energy-momentum tensor conservation we also put $\e=0$ in \eqref{deltac}). In the first case we also test what happens when we allow $\Neff$ to vary.

\subsection{Results with the full Boltzmann hierarchy for dark radiation perturbations\label{sec:fulldr}}

In our baseline case we vary the usual $\Lambda$CDM model parameters ($\ob, \oc, \theta_*, \tau, \log A_s, n_s$ as well as the 15 nuisance parameters for the TT+lowP data, plus an additional 12 for the TE+EE data), and in addition $\e$ and $f_{\dr*}$. For the definition of the standard parameters, see \cite{PCP13}. The 95\% C.L. upper bounds for the two primary dr parameters (and the derived parameter $f_{\dr0}$) are given in \tab{tab:numbers}.

If we fix the cosmological parameters other than $\e$ and $f_{\dr*}$ to their best-fit vanilla \LCDM values, the TT+lowP data alone give the constraints $\e<0.4\times 10^{-3}$, $f_{\dr*}<0.015$. These constraints slightly tighten if the nuisance parameters are also fixed (see \tab{tab:numbers}). There is no significant improvement when additional datasets are added, as they mainly serve to constrain the vanilla parameters. The CMB anisotropies are more sensitive to change in the temperature scaling than the combination of absorption line and tSZ measurements by one to two orders of magnitude, if degeneracies are not taken into account. The constraint on $\e$ is of \emph{the same order of magnitude as the one from spectral distortion} \cite{Ellis:2013}. (If it were stronger, there would be no need to to worry about tuning the model separately so as not to distort the blackbody spectrum, because not spoiling the anisotropy pattern would be even more difficult.)

\begin{figure*}[t]
\centering
\includegraphics[width=0.94\textwidth]{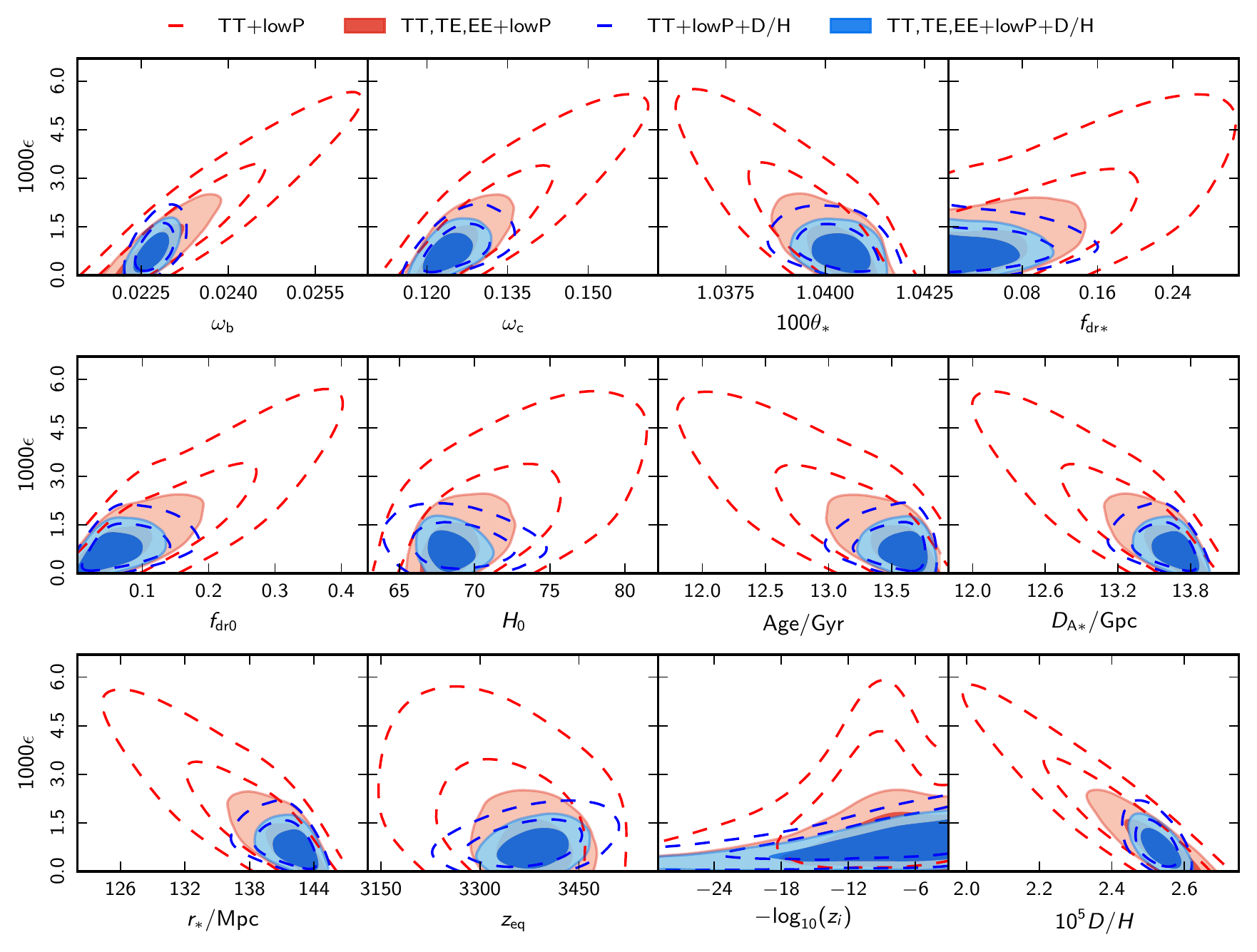}
\includegraphics[width=0.94\textwidth]{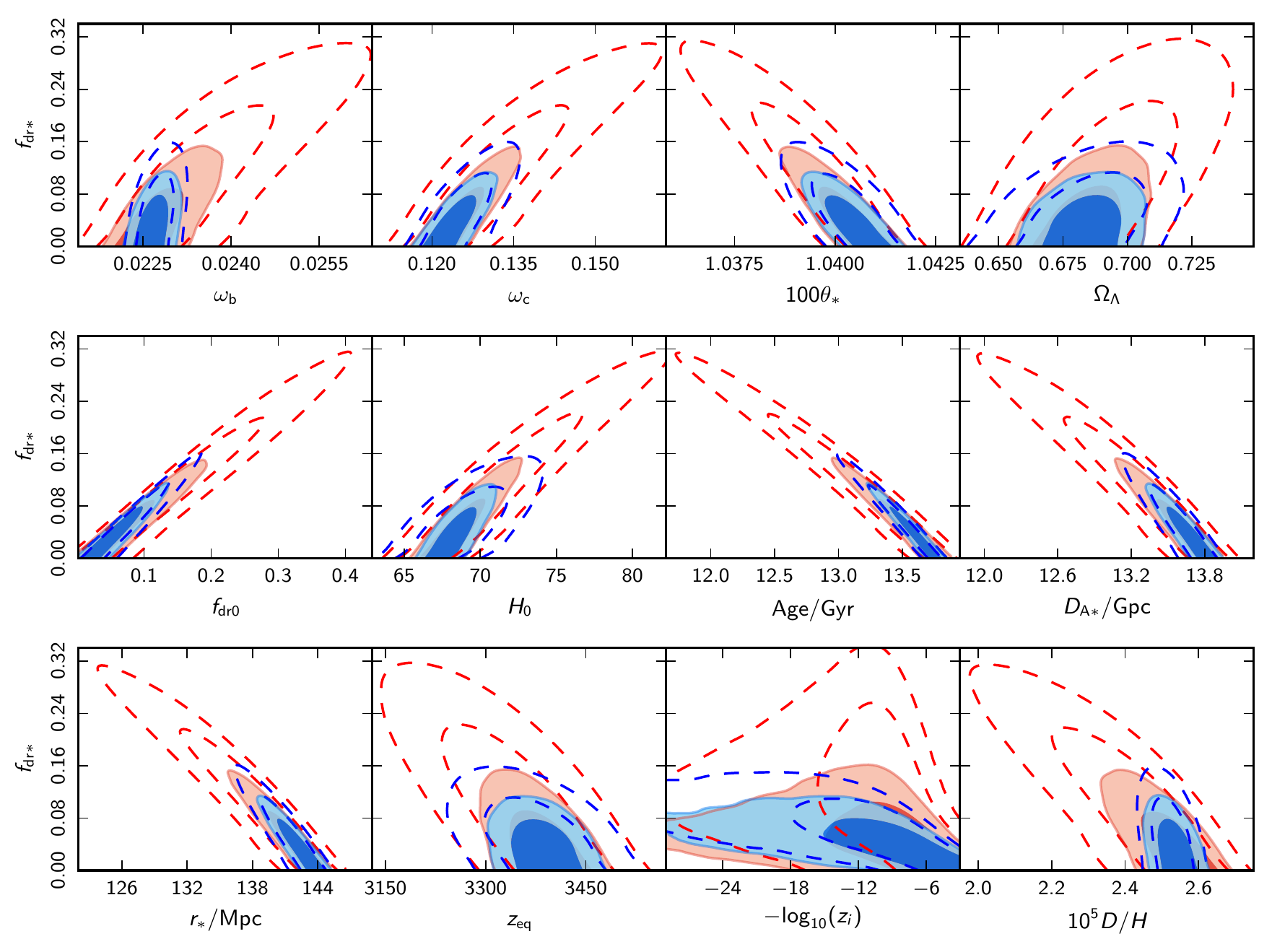}\\
\vspace{-5mm}
\caption{Degeneracies of selected parameters with $\epsilon$ (top 3 rows) and $f_{\rm dr\ast}$ (bottom 3 rows) in the full dr model, with the 68\% C.L. and 95\% C.L. posterior regions for various data sets.\label{fig:drfull_2d}
\vspace{-1cm}}
\end{figure*}

However, in contrast to spectral distortion, the effect of $\e$ on the anisotropies is degenerate with changes in other parameters; see figure \ref{fig:drfull_2d}. The limits relax by one order of magnitude when the other parameters are not artificially fixed. From the TT+lowP data alone, we get $\e<4.5\times10^{-3}$, $f_{\dr*}<0.26$. The numbers for different data combinations and treatments of the perturbations are listed in \tab{tab:numbers}.

\begin{figure}[t]
\centering
\includegraphics[width=\textwidth]{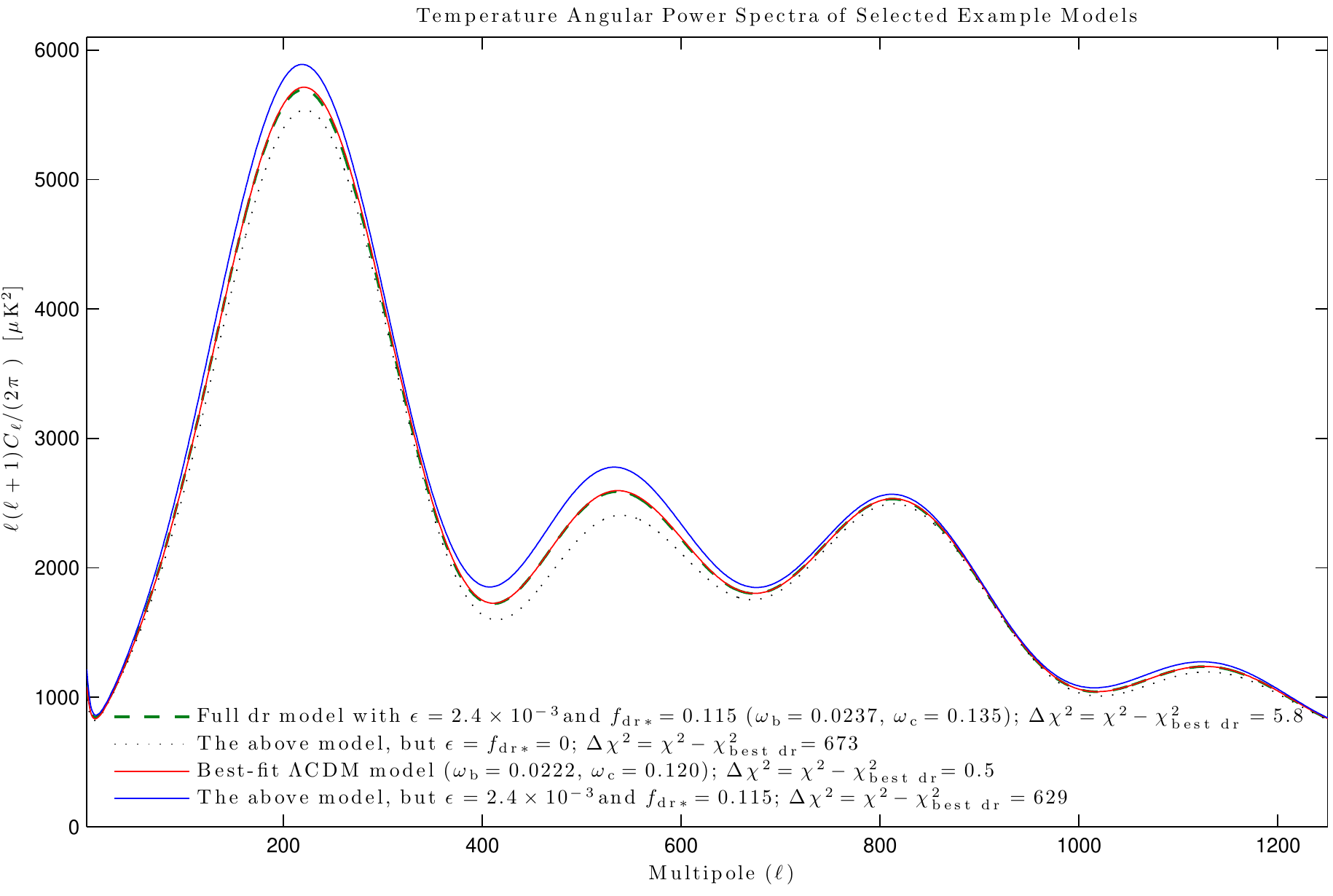}
\caption{The effect on the CMB temperature angular power spectrum. The goodness-of-fit values $\chi^2$ have been obtained with the Planck TT+lowP data. Due to cosmic variance, the lowP ($\ell=2$--$29$) data gives almost identical $\chi^2$ for all the example cases. The first model has been selected by requiring $\Delta\chi^2 < 6$, which roughly corresponds to $2\sigma$, and then finding the model with the maximum $\e$. (This model also happens to have the maximum $f_{\dr*}$.) The large values of $\ob=0.0237$ and $\oc=0.135$ almost perfectly compensate for the effect of the relatively large $\e=2.4\times10^{-3}$ and $f_{\dr*}=0.115$ ($f_{\dr0}=0.173$).}
\label{fig:Cl}
\end{figure}

\begin{figure}[t]
\includegraphics[width=\textwidth]{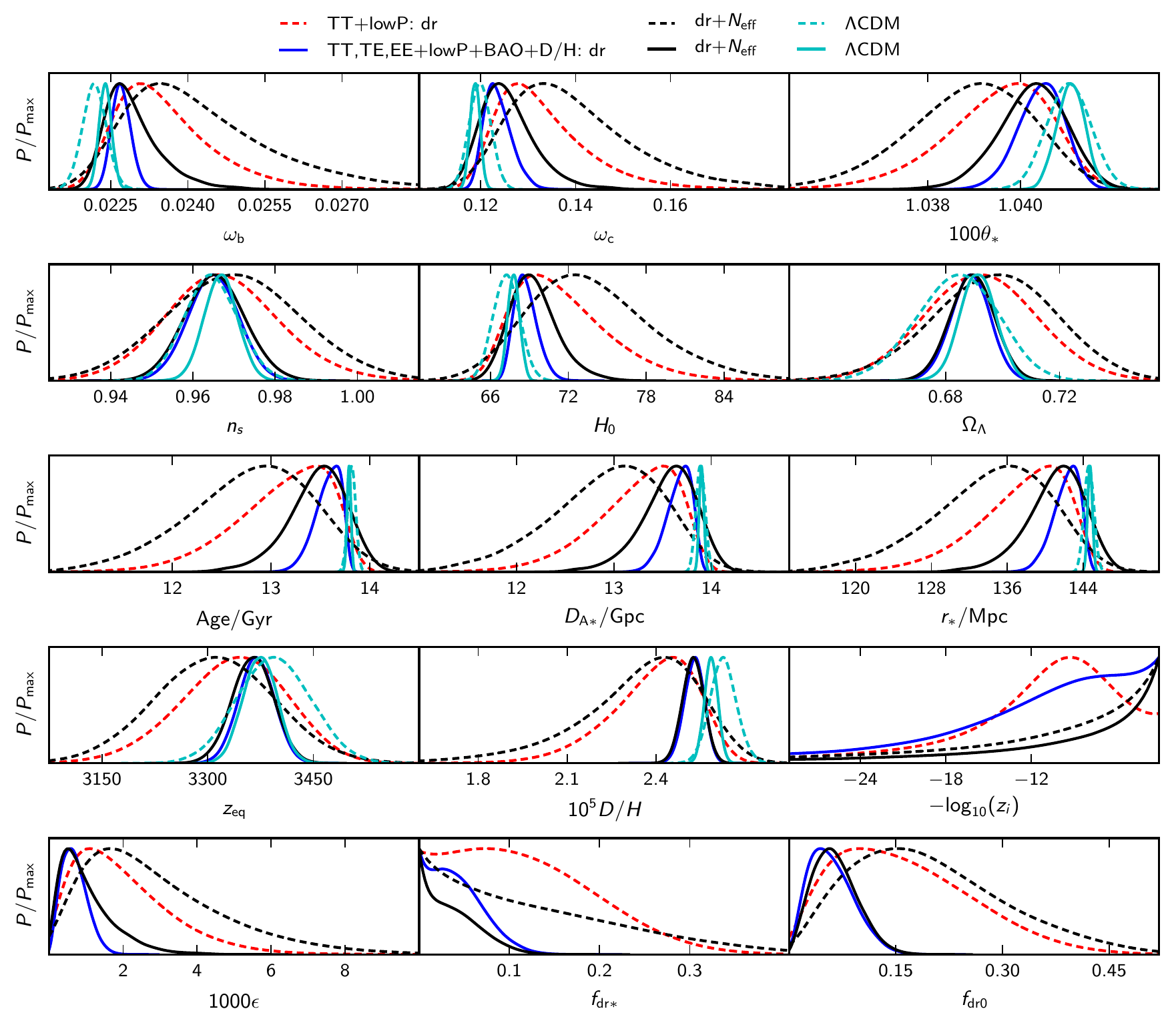}\\
\vspace{-10mm}
\caption{1d marginalised posterior probability density functions of selected parameters. \label{fig:dr_1d}}
\end{figure}

As shown in \fig{fig:drfull_2d}, the well-fitting values of $\e$ (and $f_{\dr*}$) are positively correlated with both $\ob$ and $\oc$ (and hence also with $\om$). This strong degeneracy is due to the fact that the height of the acoustic peaks in the temperature angular power spectrum are sensitive to the ratio $\bar\rho_\mathrm{b}/\bar\rho_\c$. Larger $\e$ corresponds to larger $\bar\rho_\c$ at early times, which can be compensated by increasing $\bar\rho_\mathrm{b}$. The densities $\ob$ and $\oc$ are positively correlated with each other, in contrast to the vanilla \LCDM case, where they are anticorrelated. This can be explained by the need to change $\oc$ in tune with $\ob$ so as to keep the peak height ratios fixed (see e.g. \cite{Mukhanov:2003}). In figure \ref{fig:Cl} we show how the peaks are significantly raised by the change in the temperature scaling and the inclusion of dark radiation. The change in the peak height ratios is also clear, and without changing the vanilla parameters to compensate, the $\chi^2$ is worse by more than 600 (compare the red and blue lines). Increasing $\ob$ and $\oc$ allows for a good fit even with a significant amount of dark radiation (green dashed line).

For clarity, we do not show in \fig{fig:drfull_2d} data combinations that involve BAO. The combination TT+lowP+BAO is virtually indistinguishable from TT+lowP, and likewise TT,TE,EE+lowP+BAO from TT,TE,EE+lowP. This is due to the fact that BAO data are mostly sensitive to $\Omega_\Lambda$ (or $\Om = 1 - \Omega_\Lambda$, not $\om$), but adding BAO data does not significantly shift $\Omega_\Lambda$. Thus, the only noticeable effect is slightly tighter constraints on $n_s$, $H_0$ and $\Omega_\Lambda$. In contrast, the D/H prior constrains $\ob$ directly, reducing the ($\ob$,$\,\e$) and ($\ob$,$\,f_{\dr*}$) degeneracies significantly and improving the limits to $\e<1.8\times10^{-3}$, $f_{\dr*}<0.13$. Adding the high-multipole TE,EE data also reduces the degeneracy by improving the constraints on standard parameters. Hence, we can most efficiently reduce the degeneracies by using TT,TE,EE+lowP+D/H. Comparison of the third last and last columns of \tab{tab:numbers} shows that adding BAO to this combination marginally tightens the constraint on $\e$, but actually weakens the constraint on $f_{\dr*}$. Since we are mainly interested in $\e$, we quote here and in the abstract as our definite constraints the numbers with the maximal data set TT,TE,EE+lowP+BAO+D/H (the last column): $\e<1.3\times10^{-3}$, $f_{\dr*}<0.095$.

As the sound horizon at last scattering $r_*$ is proportional to $\ob^{-1/2}$ and $\om^{-1/2}$, it is correspondingly negatively correlated with $\e$ (see appendix \ref{sec:standard}). The angle covered by the sound horizon remains precisely determined (at the few per mille level, to the value $\theta_* = 1.040\times 10^{-2}\,$rad), so the angular diameter distance $D_{A*}=r_*/\theta_*$ has the same negative correlation with $\e$ as $r_*$. Because $D_{A*}\propto H_0^{-1}$, the Hubble constant $H_0$ is shifted upwards, from $H_0 = 67.2\pm1.0\,$km$\,$s$^{-1}$Mpc$^{-1}$ to $H_0=71.6^{+2.3}_{-4.5}\,$km$\,$s$^{-1}$Mpc$^{-1}$, and the age of the universe is correspondingly shifted downwards, from the vanilla \LCDM value $t_0 = 13.82\pm0.04\,$Gyr to  $t_0=13.1^{+0.6}_{-0.3}\,$Gyr. When more data are added, the shift becomes smaller.

The 1d marginalised posterior probability densities for selected parameters with our minimal and maximal datasets are shown in \fig{fig:dr_1d}. For comparison we also indicate the vanilla \LCDM results.  While the standard cosmological parameters can shift by several standard deviations in terms of the vanilla model error bars (more than 18$\sigma$ in the case of $t_0$), the shifts are not more than 2$\sigma$ in terms of the dr model error bars. In the dr model some 1d posteriors are noticeably non-Gaussian, in particular those of $\ob$, $\oc$, $H_0$, $t_0$, $r_*$, and $D_{A*}$. As the goodness-of-fit (best-fit $\chi^2$) is not improved by the addition of dr, the parameter combinations with standard parameters at almost their best-fit vanilla \LCDM values and $\e$ and $f_{\dr*}$ almost zero receive quite a large weight. However, as explained above and seen in \fig{fig:Cl}, with larger $\ob$ and $\oc$ one can compensate the changes caused by non-zero $\e$ and $f_{\dr*}$. This results in a long tail at large $\ob$ and $\oc$, reflected as a long tail at small values of $r_*$, $D_{A*}$, and $t_0$ and large values of $H_0$. Adding high-multipole TE+EE data and/or D/H prior significantly reduces these tails.

\begin{figure*}[t]
\includegraphics[width=0.5\textwidth]{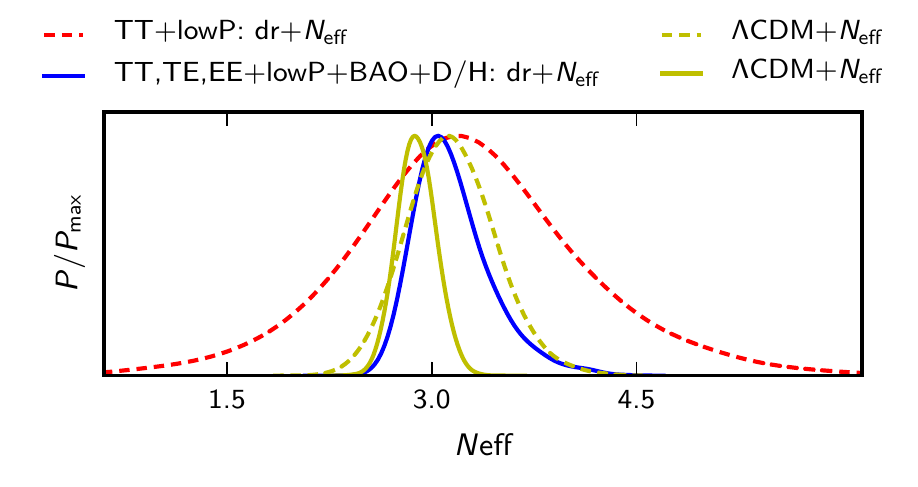}
\includegraphics[width=0.5\textwidth]{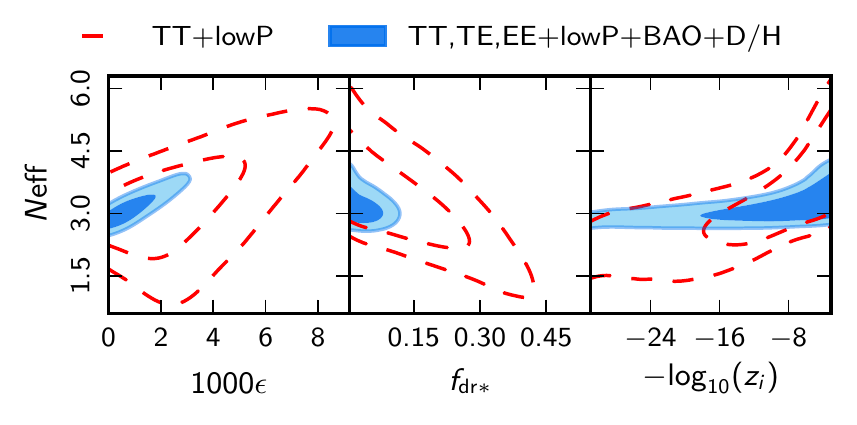}\\
\vspace{-10mm}
\caption{Left panel: 1d marginalised posterior probability density of $N_{\rm eff}$.  Right panels: degeneracy between dr parameters and $N_{\rm eff}$. \label{fig:Neff}}
\end{figure*}
\begin{figure}[t]
\centering
\includegraphics[width=0.94\textwidth]{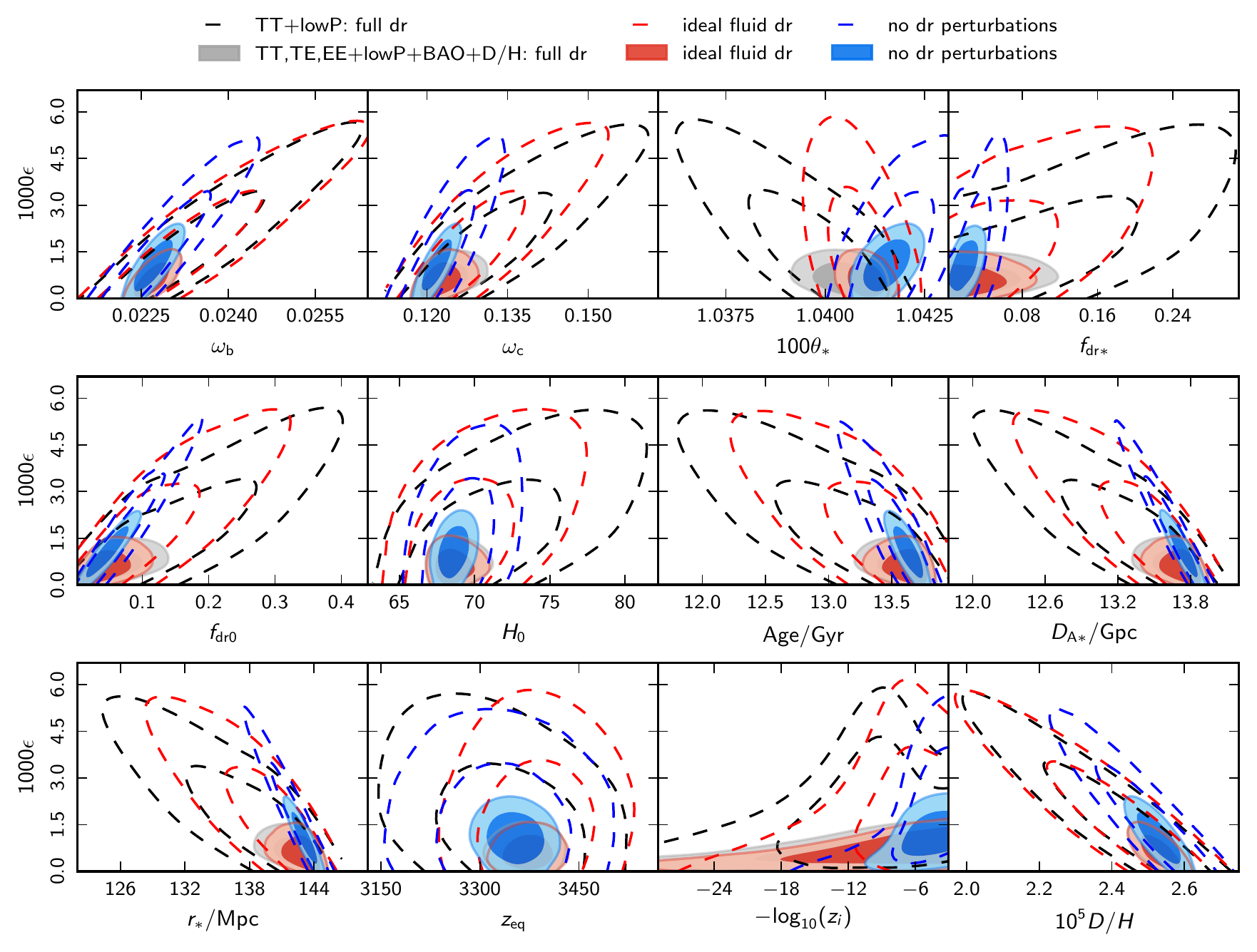}
\includegraphics[width=0.94\textwidth]{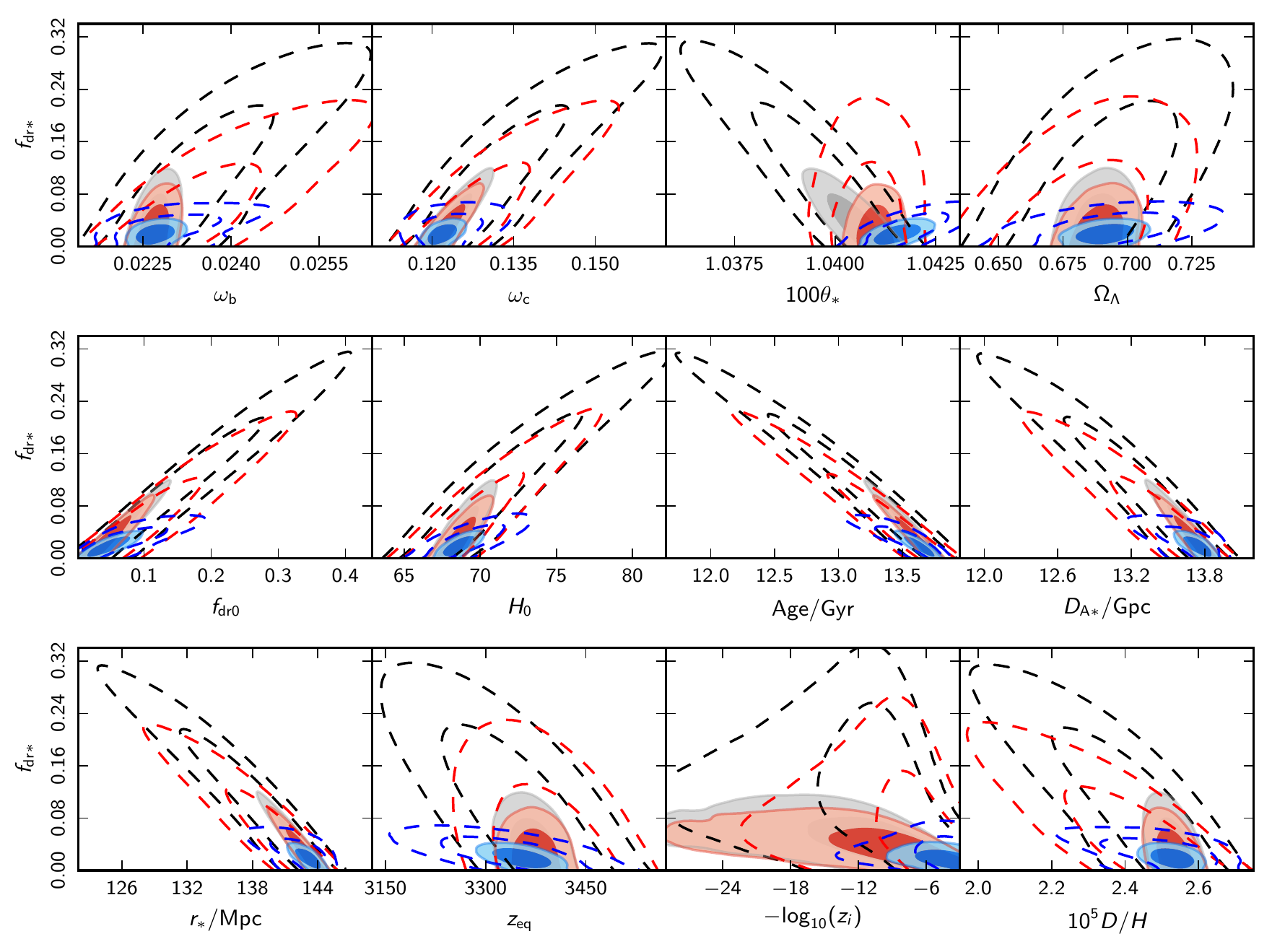}\\
\vspace{-5mm}
\caption{Comparison of degeneracies of parameters with $\epsilon$ (top 3 rows) and with $f_{\rm dr\ast}$ (bottom 3 rows) in three different treatments of dr perturbations with our minimal and maximal datasets.\label{fig:drideal_2d}
\vspace{-1cm}
}
\end{figure}

\subsection{Allowing the number of relativistic degrees of freedom $\Neff$ to vary \label{sec:Neff}}

We also consider a case where the number of relativistic degrees of freedom $\Neff$ other than photons and dr is not fixed to the standard value $3.046$, but is left as a free parameter. (The value could be higher if there are new light degrees of freedom, and smaller if there are late decaying particles that contribute more to photons than neutrinos \cite{Rossi:2014, deSalas:2015}.)  The dr constraints weaken by less than a factor of 1.5. The range of $\Neff$ changes from the vanilla $\Lambda$CDM+$\Neff$ result $3.01<\Neff<3.20$ to the dr+$\Neff$ result $1.57<\Neff<5.05$ for TT+lowP, and from $2.62<\Neff<3.17$ to $2.64<\Neff<3.77$
for the maximal dataset, as shown in \fig{fig:Neff}. The fraction $f_{\dr*}$ is negatively correlated with $\Neff$, as their effect is similar. However, the correlation between $\e$ and $\Neff$ is instead positive. This is somewhat surprising, especially as $\e$ and $f_{\dr*}$ are positively correlated for the TT+lowP dataset, and for extended datasets their correlation is small (see the fourth panel of \fig{fig:drfull_2d}). The reason may be that for a fixed $f_{\dr*}$, a larger $\e$ decreases the impact of dr at times before decoupling. This interpretation is supported by the last panel of \fig{fig:Neff}, which indicates that the larger the value of $\Neff$ the later (closer to last scattering surface) the photon-dr coupling is allowed to be turned on; compare also $-\log_{10}z_i$ in the dr and dr+$\Neff$ cases in \fig{fig:dr_1d}.

There are constraints on $\Neff$ from big bang nucleosynthesis (BBN), but because changing the temperature-redshift relation also affects nucleosynthesis, they cannot be translated into constraints on $f_{\dr*}$ or $\e$ without redoing the BBN analysis \cite{Freese:1987}. In fact, for $z_{i}>10^8$ we should in principle take into account the effect of the dr on BBN, as it could change the hydrogen and helium abundances at decoupling. However, we expect that this would have only a small effect on our results, as the amount of dr at early times is typically small in comparison to the usually allowed range of $\Neff$.

\subsection{Treating dark radiation as an ideal fluid or neglecting the dr perturbations\label{sec:ideal}}

Considering dr as an ideal fluid, \ie neglecting anisotropic stress and higher moments in the Boltzmann hierarchy, does not have a large effect.
In the ideal fluid case, the constraints on $\e$ are almost unchanged, but the constraint on $f_{\dr*}$ tightens slightly to $f_{\dr*}<0.18$ ($f_{\dr*}<0.08$ for the maximal dataset), as listed in \tab{tab:numbers}.
It might have been expected that dropping the anisotropic stress would have made both constraints weaker, on the grounds that if the data do not prefer dr, then decreasing its impact on the physics would be favoured, but this is not the case.

Neglecting the dr perturbations altogether has a larger effect, and again the impact on $f_{\dr*}$ is more pronounced: the constraints become tighter, $f_{\dr*}<0.06$ ($f_{\dr*}<0.04$ for the maximal dataset).
In other words, if adding dr to the background makes the fit worse, then its perturbations act in the opposite direction, weakening the constraints. (Note that the perturbative description does not involve any additional parameters.)
Also, the correlation of $\e$ and $f_{\dr*}$ with many standard parameters becomes weaker when perturbations are ignored, and the ideal fluid case lies between the full case and the background-only treatment, as shown in \fig{fig:drideal_2d}. Regardless of the treatment of perturbations, the goodness-of-fit is not improved by the presence of dr.

\subsection{Comparison to existing CMB constraints; allowing $\epsilon<0$ \label{sec:Planck}}

The Planck analysis \cite{Planck:cosmo}, which considered only
a change in the background, obtained the constraint
$\e=(-0.2\pm1.4)\times10^{-3}$ ($-3.0\times10^{-3}<\e<2.6\times10^{-3}$) for TT+lowP+BAO data, and
$\e=(-0.4\pm1.1)\times10^{-3}$ ($-2.6\times10^{-3}<\e<1.8\times10^{-3}$) for TT,TE,EE+lowP+BAO data. In the Planck analysis, $\e$ was allowed to have either sign.
It was also different in that there was no dr, in which case $\e\neq0$ also affects the scaling of the total background radiation energy density.
In \tab{tab:numbersNegEps} we compare our results to those of the Planck collaboration. In order to make the models more comparable, we also consider our model in the case where negative values of $\epsilon$ are allowed.

When $\epsilon$ is negative, we need to make sure that the dr density today remains positive. Therefore, according to \eqref{eq:rhodr0}, we introduce a prior requirement $(1- f_{\dr*}) a_{{\mathrm{ref}}}^{4\e} < 1$ between $\epsilon$ and $f_{\dr*}$. In addition, with $\epsilon<0$, we always set $z_i=10^{29}$. The first condition reduces the prior parameter space volume of the negative $\epsilon$ models compared to the positive $\epsilon$ case. The second condition disfavours large negative values of $\epsilon$, since the interaction is turned on very early, compared to the positive $\epsilon$ models, where $z_i$ is sometimes much closer to last scattering.
Despite the large difference in the models, in all cases our constraints on $\epsilon$ are of the same order of magnitude as those obtained by the Planck collaboration, as seen in \mbox{\tab{tab:numbersNegEps}}.


\setlength{\tabcolsep}{.33667em}
\begin{table}[b]
\footnotesize
\centering
\begin{tabular}{l|rr|rr}
\hline
\hline
                                        &  \multicolumn{2}{l|}{TT+lowP} & \multicolumn{2}{l}{TT,TE,EE+lowP}  \\
                                        &  +BAO  & +BAO+D/H    &  +BAO    & +BAO+D/H   \\
\hline
$1000\epsilon\!:\ \ $ Full dr ($\epsilon \ge 0$; our main case) &   $(0;\ 4.4)$ &  $(0;\ 1.6)$ &  $(0;\ 2.0)$ &  $(0;\ 1.3)$  \\
$\phantom{1000\epsilon\!:\ \ }$ Full dr ($\epsilon<0$ allowed) &  $( -1.4;\   4.5)$ &   $( -0.4;\   1.6)$ &   $( -1.0;\   2.0)$ &   $( -0.3;\   1.4)$  \\
$\phantom{1000\epsilon\!:\ \ }$ No dr perturbations ($\epsilon<0$ allowed) &  $( -0.7;\   4.4)$ &   $(  0.1;\   2.6)$ &   $( -0.4;\   3.5)$ &   $(  0.1;\   2.1)$  \\
$\phantom{1000\epsilon\!:\ \ }$ Planck collaboration \cite{Planck:cosmo}; no dr,   &                            &                            &                            &                             \\
$\phantom{1000\epsilon\!:\ \ }$ background $\bar\rho_\c\propto (1+z)^{4(1+\e)}$ & $(-3.0;\ 2.6)$ & & $(-2.6;\ 1.8)$ & \\
\hline
\hline
\end{tabular}
\caption{\label{tab:numbersNegEps}The posterior 95\% C.L. intervals for $1000\epsilon$. The first line repeats our main results with $\epsilon \ge 0$ from \tab{tab:numbers}. On the next two lines we report constraints in our model when we allow negative values of $\epsilon$. The last line lists the constraints obtained by the Planck collaboration for a different model, where only the background scaling of $\rho_\c$ was modified \cite{Planck:cosmo}.
}
\end{table}

\section{Conclusions} \label{sec:conc}

\para{Tight constraints on temperature scaling.}

We have tested deviations from the standard temperature-redshift relation of the form $T\propto(1+z)^{1+\e}$ using CMB anisotropies, with changes to the perturbation equations taken into account for the first time. The Planck collaboration considered such a test, but looked only at the effect of the change of the background \cite{Planck:cosmo}. Assuming a blackbody spectrum, such a change is equivalent to modifying the distance duality relation as $D_L=(1+z)^{2(1+\e)}D_A$. We have used the CMB anisotropies as measured by the second data release of Planck (published in the Planck Legacy Archive in July 2015), supplemented by BAO data \cite{BAO} and measurements of deuterium abundance, D/H=$(2.53\pm0.04)\times10^{-5}$ \cite{Cooke:2013}.
Such a test presents a major increase in the redshift range compared to constraints from absorption lines and the tSZ effect, which go up to $z=3.025$. In a sense, the test extends even earlier than the last scattering surface at $z=1090$, because a modified scaling relation changes the evolution of the photon density at early times, which has an impact on the dynamics at last scattering.

Including perturbations makes the treatment more model-dependent. We have simply added a species of dark radiation (dr) particles, into which photon energy density is transferred, to obtain a consistent set of equations. The total energy density of photons plus dr particles evolves as in the usual case ($\propto a^{-4}$), so our dr is not directly limited by the constraints on the number of extra relativistic degrees of freedom $\Neff$ \cite{Rossi:2014}.
The model also involves the parameter $f_{\dr*}$, which measures the ratio of the dr energy density to the sum of dr and photon energy densities around last scattering.
Compared to the observations of absorption lines and the tSZ effect, the CMB anisotropies strengthen the limit on $\e$ by an order of magnitude, from $-24\times10^{-3}<\e<11\times10^{-3}$ to $\e<1.8\times10^{-3}$ for TT+lowP+D/H and $\e<1.3\times10^{-3}$ for our maximal dataset TT,TE,EE+lowP+BAO+D/H. The corresponding limits on $f_{\dr*}$ are $f_{\dr*}<0.128$ and $f_{\dr*}<0.095$. Details of the constraints for different data combinations are given in \tab{tab:numbers}.

In addition to calculating the dr Boltzmann hierarchy self-consistently, we have considered two alternative treatments: truncating at the ideal fluid level or neglecting dr perturbations altogether. Treating dr as an ideal fluid makes negligible difference on $\e$ and a moderate difference on $f_{\dr*}$, while dropping perturbations altogether loosens the upper limit on $\e$ to $2.5\times10^{-3}$, but tightens the limit on $f_{\dr*}$ to $0.050$ for TT+lowP+D/H. For the maximal dataset we find without dr perturbations $\e<2.1\times10^{-3}$ and $f_{\dr*}<0.036$. In other words, including perturbations does not have a large impact on $\e$, but (unlike one might naively expect) ignoring them may lead to artificially tight constraints on $f_{\dr*}$. Another model with the same background $T(z)$ relation could have very different perturbation dynamics, leading to stronger or weaker constraints. However, our results show that it is at least possible to construct a consistent perturbative model where considering only the background captures the effect of $\epsilon$ on the CMB anisotropies, to within a factor of two, providing an estimate of the reliability of the analysis of the Planck collaboration, where perturbations were not taken into account \cite{Planck:cosmo}.

\para{The role of degeneracies.}

The CMB anisotropies are highly sensitive to changing the temperature-redshift scaling, and the strength of the constraints is mainly limited by degeneracies: $\e$ is particularly degenerate with the physical baryon density $\ob$ and the physical cold dark matter density $\oc$. If we fixed these and other cosmological parameters to their best-fit vanilla \LCDM values, the constraint on $\e$ would be an order of magnitude stronger. Such a constraint would be competitive with the one from distortion of the blackbody spectrum, which currently puts the strongest limit on possible changes to the temperature-redshift relation \cite{Ellis:2013}, though it is possible to avoid it by suitably tuning the change in photon energy and number density \cite{Avgoustidis:2011, Chluba:2014}.

In order to improve the constraints obtained from the CMB anisotropies, independent measurements of some of the cosmological parameters that are most degenerate with $\e$ and $f_{\dr*}$ are required. We have seen that a prior on D/H reduces the $\ob$ degeneracy significantly. An independent measurement of a low Hubble parameter, $H_0 \lesssim 69\,$km$\,$s$^{-1}$Mpc$^{-1}$, would also further constrain dark radiation.

\acknowledgments
The CMB temperature and polarisation likelihoods used in this work are described in \cite{Aghanim:2015xee} and available via Planck Legacy Archive. They are based on observations obtained with Planck, an ESA science mission with instruments and contributions directly funded by ESA Member States, NASA, and Canada. JV and VK were supported by the Academy of Finland grant 257989.  We thank CSC --- IT Center for Science Ltd (Finland) for computational resources. Part of the results were achieved using the PRACE-3IP project (FP7 RI-312763) resource Sisu at CSC.

\appendix

\section{Modified sound horizon \label{sec:standard}}

The comoving sound horizon at redshift $z$ is
\bea
r(z) = \int_0^{\eta(z)} c_s(\tilde\eta) d\tilde\eta\,,
\label{eq:rs}   
\eea
where $c_s$ is the sound speed in the photon-baryon fluid. The sound horizon is the distance a sound wave has been able to propagate since the big bang. The sound speed is given by
\bea
c_s^2 = \frac{\delta p}{\delta\rho} = \frac{\delta p_\gamma}{\delta\rho_\gamma + \delta\rho_{\rm b}} = \frac{\dot{\bar p}_\gamma}{\dot{\bar\rho}_\gamma + \dot{\bar\rho}_{\rm b}} = \frac{1}{3}\frac{\dot{\bar \rho}_\gamma}{\dot{\bar\rho}_\gamma + \dot{\bar\rho}_{\rm b}}\,.
\eea
In the standard vanilla \LCDM calculation we would replace $\dot{\bar\rho}_{\rm b}$ by $-3H\bar\rho_{\rm b}$ and $\dot{\bar\rho}_\gamma$ by  $-4H\bar\rho_\gamma$, according to the continuity equations. However, in our model the continuity equation for photons \eqref{rhogamma1} contains $\e$. Hence, after using the continuity equations, the result is
\bea
c_s^2 = \frac{1}{3} \frac{1}{1+\frac{R}{1+\e}}\,, \mbox{ where } R(\eta)= \frac{3}{4} \frac{\bar\rho_{\rm b}(\eta)}{\bar\rho_\gamma(\eta)}\,.
\label{eq:cs2}
\eea
In the case $\e=0$ this reduces to the vanilla \LCDM result. In our model, there is another $\e$ dependency hidden in the baryon-to-photon ratio $R$, namely
\bea
R = \frac{3}{4} \frac{\bar\rho_{\rm b0} a^{-3}}{\bar\rho_{\gamma0} a^{-4(1+\e)}} =  \frac{3}{4} \frac{\bar\rho_{\rm b0}}{\bar\rho_{\gamma0}} a^{1+4\e} =  \frac{3}{4}\frac{\ob}{\omega_\gamma} a^{1+4\e} = R_0 a \times a^{4\e}\,,
\label{eq:R}
\eea
where today's baryon-to-photon ratio is $R_0 = \frac{3}{4}\bar\rho_{\rm b0}/\bar\rho_{\gamma0} = \frac{3}{4}\ob/\omega_\gamma$.  With $\bar\rho_{\c0} = \textstyle \frac{\pi^2}{15} \bar T_0^4$ and $\bar T_0=2.7255\,$K \cite{Fixsen:2009ug}, we obtain $\omega_\gamma = 2.4728\times10^{-5}$. Substituting $R$ from \eqref{eq:R} to \eqref{eq:cs2} and this further to \eqref{eq:rs} leads to
\bea
r(z) = \frac{1}{\sqrt{3}}\int_0^{a(z)} \frac{1}{\sqrt{1+ \frac{R_0 a^{1+4\e}}{1+\e}}} \frac{d\eta}{da} da\,.
\eea
Here $d\eta / da = \dot a^{-1}$, where $\dot a$ is given by the Friedmann equation
\bea
\dot a = H_0 a^2 \sqrt{\Omega_{\rm r,tot}a^{-4} + \Om a^{-3} + \Omega_\Lambda} \approx H_0 \sqrt{\Omega_{\rm r,tot} + \Om a} = H_0\sqrt{\Om} \sqrt{a_{\rm eq}  + a}\,,
\eea
where $a_{\rm eq} = \Omega_{\rm r,tot} / \Om = \omega_{\rm r} / \om$. As the range of interest is $0 < a \lesssim 10^{-3}$, we can ignore the $\Omega_\Lambda$ term.  Since $H_0^{-1} \approx h^{-1}2998\,$Mpc, we find
\bea
r(z) & = & \!\frac{2998\,\rm Mpc}{\sqrt{3\om}} \int_0^{a(z)} \frac{1}{\sqrt{1+ \frac{R_0 a^{1+4\e}}{1+\e}}} \frac{1}{\sqrt{a_{\rm eq}+a}} da\\
 & =  & \! \frac{2}{3} \frac{2998\,\rm Mpc}{\sqrt{\om}} \frac{\sqrt{(1+\e)\omega_\gamma}}{\sqrt{\ob}} \int_0^{a(z)} \!\!\left[a^{2+4\e} + a_{\rm eq}  a^{1+4\e} +\frac{1+\e}{R_0} a + \frac{1+\e}{R_0}a_{\rm eq}\right]^{-\half}\!\!\!\!da\,.
\label{eq:final_rs}
\eea
On the second line we have taken $\sqrt{(1+\e)/R_0}$ out from the integral and substituted $R_0$ defined after \eqref{eq:R}. In the vanilla case the remaining integral is easy to calculate analytically, and it is slightly less than unity. With $\e\neq0$, there is no analytical solution. We have edited \texttt{CAMB} to use \eqref{eq:cs2} with $R$ given by \eqref{eq:R} when integrating the sound speed according to \eqref{eq:rs} to find $r_{\rm drag}$ (needed when using the BAO data) and $r_*$, and when determining the derived parameter $H_0$ from the primary parameter $\theta_* = r_* / D_{A*}$ drawn by \texttt{CosmoMC}.

Result \eqref{eq:final_rs} is important in explaining why $r_*$ and $r_{\rm drag}$ are negatively correlated with $\e$. Although there is an intrinsic positive correlation from the $\sqrt{1+\e}$ factor and $\e$ terms in the integral, the dominant dependence is $\propto(\om\ob)^{-1/2}$, since $\e\ll1$. As explained in the main text, both $\om$ and $\ob$ are positively correlated with $\e$ in well-fitting models, so $r$ is negatively correlated with $\e$.

\providecommand{\href}[2]{#2}\begingroup\raggedright

\end{document}